# Giant Photoresponsivity of Mid-Infrared Hyperbolic Metamaterials in the Quantum Regime

Pai-Yen Chen[1,*], Mehdi Hajizadegan[1], Maryam Sakhdari[1], and Andrea Alù[2]

[1]Department of Electrical and Computer Engineering, Wayne State University, Detroit, MI, 48202, U.S.A.

[2]Department of Electrical and Computer Engineering, University of Texas at Austin, Austin, TX, 78712, U.S.A.

*pychen@wayne.edu

*We explore broadband, wide-angle mid-infrared rectification based on nanopatterned hyperbolic metamaterials (HMMs), composed of two dissimilar metals separated by a sub-nanometer tunnel barrier. The exotic slow-light modes supported by such periodically trenched HMMs efficiently trap incident radiation in massively parallel metal-insulator-metal tunnel junctions supporting ultrafast optical rectification induced by photon-assisted tunneling. This leads to highly efficient photon-to-electron conversion, orders of magnitude larger than conventional optical rectennas. Our results promise an impact on infrared energy harvesters and plasmonic photodetectors.*

Photon-assisted tunneling is an intrinsic quantum-size effect in plasmonic nanostructures, which has recently attracted growing interest because of its exotic nonlinear and nonlocal optical properties [1]-[7]. The excitation of surface plasmon polariton resonances, combined with tunneling-induced optical nonlinearities, may allow a rich variety of nonlinear optical phenomena, such as efficient harmonic generation, multi-wave mixing, resistive switching, and two-photon absorption [7]-[11]. The second-order phenomenon of optical rectification is of particular interest, since it may find potential applications in photodetection and photovoltaics at mid-infrared (MIR) wavelengths [12]-[13]. In the MIR regime, a major thrust of renewable energy research contains harvesting the roughly $10^{17}$ W of infrared thermal radiation that Earth continuously emits into outer space [14]. Nonetheless, there exists no suitable MIR-bandgap semiconductors [14]-[15] that can make an efficient interband photodetector or photovoltaic device, due to the frequent Auger recombination and generation at room temperature. Arguably, rectifying antennas (or "rectennas"), which have been commonly used to generate power from microwaves with a > 90% power conversion efficiency [16], may realize emissive energy harvesters [12]-[14]. In this context, a nano-rectenna operated at infrared wavelengths can collect infrared radiation and, when loaded with a MIM tunneling diode, can convert light into electricity through nonlinear optical rectification [12]-[14]. However, to date, proof-of-principle nano-rectennas have shown insufficient responsivity (or external quantum efficiency) at low bias voltages (less than mA/W) [17]-[19], far from practical use in photodetection (~ 0.1 A/W) and photovoltaics. The major roadblock toward high-performance and viable infrared rectennas stems from inefficient light coupling, yielded by a tradeoff between maximum parasitic cutoff frequency and device dimensions, i.e., diode area and insulator (tunnel barrier) thickness [12], an issue that remains unsolved in conventional antenna-diode architectures.

In this Letter, we explore the possibility of using a metamaterial approach, namely hyperbolic metamaterials (HMMs) formed by parallel arrays of asymmetric MIM tunnel junctions, to realize highly efficient mid-infrared rectification, as shown in Fig. 1(a). In this multilayered HMM structure, the effective permittivity tensor is $\bar{\bar{\varepsilon}}_{eff} = \varepsilon_0(\varepsilon_\parallel \hat{x}\hat{x} + \varepsilon_\parallel \hat{y}\hat{y} + \varepsilon_\perp \hat{z}\hat{z})$ [20]-[22], where $\varepsilon_\parallel = \sum_{i=1}^{N} \varepsilon_i \rho_i$ and $\varepsilon_\perp = \left[\sum_{i=1}^{N} \varepsilon_i^{-1} \rho_i\right]^{-1}$, $\varepsilon_i$ and $\rho_i$ are relative permittivity and volume fraction of $i$-th constituent materials, respectively. A multilayered HMM can behave as a uniaxially anisotropic medium [20]-[22], provided that extreme material properties $\mathrm{Re}[\varepsilon_\perp] \cdot \mathrm{Re}[\varepsilon_{//}] < 0$ can be supported, such that the isofrequency surfaces become open hyperboloids, as illustrated in the inset of Fig. 1(a).

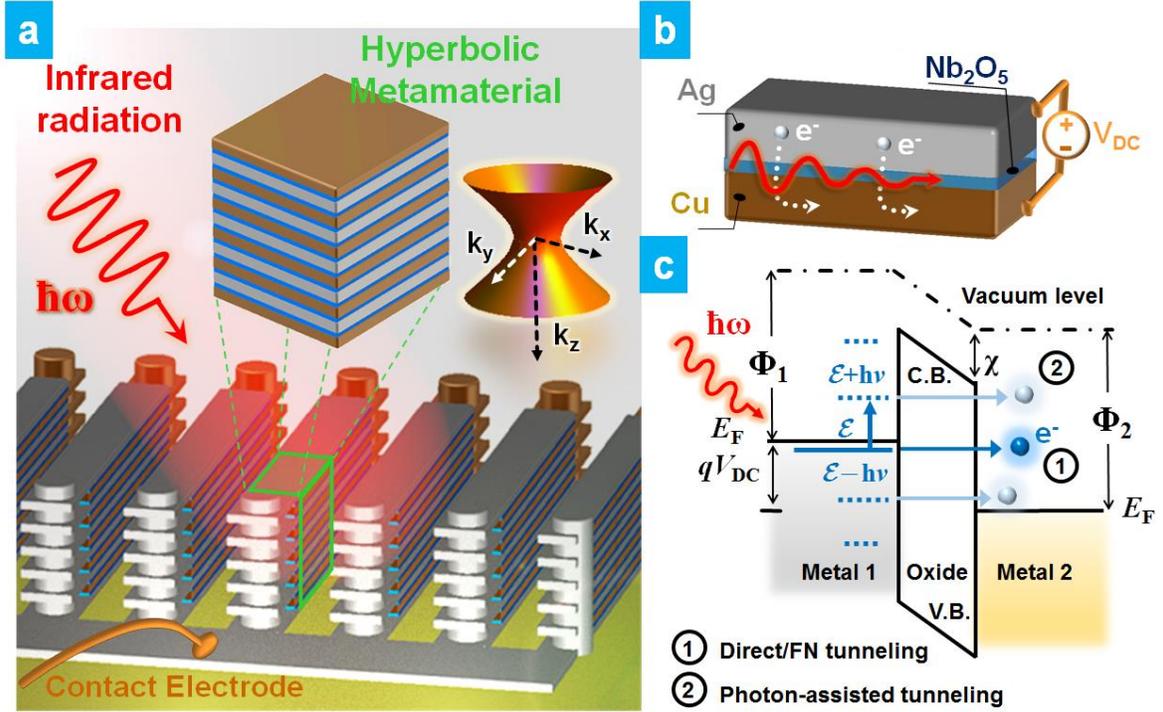

*Figure 1* *Schematics of (a) an infrared rectifying device composed of nanopatterned HMMs, (b) nonlinear optical rectification in Ag-Nb$_2$O$_5$-Cu tunnel junction forming the HMM, (c) energy band diagram of photon-assisted tunneling, (static) direct tunneling, and Fowler-Nordheim (FN) tunneling in a dissimilar MIM tunnel junction.*

So far, HMMs have been demonstrated to exhibit negative refraction, hyperlensing, Dyakonov plasmons, and anomalously large density of photonic states (PDOS) in a broad spectral range [20]-[22]. In this context, HMMs may represent an exciting platform, which holds the greatest promise for developing novel nanophotonic and quantum-optical devices, such as a non-resonant single photon sources [23] and substrates that can mold spontaneous emission into directional beams [24]. Moreover, the use of nanostructured HMM substrates (e.g., tapered HMM-waveguide array [25]) may realize "rainbow trapping", with a broadband omnidirectional optical absorption, due to the superposition of multiple slow-wave modes. Beyond these concepts, we propose here a new interesting application of nanopatterned HMM substrates combined with arras of asymmetric MIM tunneling diodes [Figs. 1(a)], aimed at achieving highly efficient mid-infrared rectification. The rainbow trapping effect in such structures may allow scattering light into plasmonic (tunneling) nanojunctions, and confine it such that the strong optical fields can substantially enhance nonlinear optical rectification.

In the following, we theoretically investigate the performance of the proposed HMM-based rectifying devices, considering the interaction of electromagnetic radiation and nonlinear quantum transport. Figure 1(c) illustrates the photon-assisted tunneling and static tunneling processes in a MIM tunnel junction: in a general scenario, the voltage across the MIM junction can be assumed to be a combination of optical and dc components $V(t) = V_{dc} + (1/2)(V_\omega e^{-i\omega t} + c.c)$. From the Tien-Gordon theory [10]-[11], the new time-dependent wavefunction for electrons in metal becomes: $\psi(\bar{\mathbf{r}},t) = \psi_0(\bar{\mathbf{r}},t)\exp\left[-i/\hbar \int^t qV(t')dt'\right]$

$= \psi_0(\bar{\mathbf{r}},t) \sum_{n=-\infty}^{n=+\infty} J_n(qV_\omega/\hbar\omega)e^{-in\omega t}$, where $q$ is the electron charge, $\hbar$ is the reduced Plank constant, $\psi_0(\bar{\mathbf{r}},t)$ is the unperturbed Schrödinger wavefunction [10]-[11], and $J_n(\cdot)$ is the $n$-th order

Bessel function of the first kind. The modified wavefunction implies that optical signals adiabatically modulate the electron potential energy. The monochromatic electromagnetic field may therefore excite new quantum-well virtual states separated from the unperturbed ground state by $\pm n\hbar\omega$, where $n$ corresponds to the number of photons absorbed or emitted, with a probability $J_n^2(qV_\omega/\hbar\omega)$, by an electron on the metal surface. As a result, the time-dependent current density is in the form of a Fourier series $J(t) = \sum_{m=0}^{m=+\infty} \frac{1}{2}\left(J_{m\omega}e^{-im\omega t} + c.c.\right)$. The dc ($m = 0$) and $m$-th frequency-dependent currents under illumination are given by

$$J_{dc} = \sum_{n=-\infty}^{\infty} J_0^2(qV_\omega/\hbar\omega) J_{dark}(qV_{dc} + n\hbar\omega) \tag{1a}$$

$$J_{m\omega} = \sum_{n=-\infty}^{\infty} J_n(qV_\omega/\hbar\omega)\left[J_{n+m}(qV_\omega/\hbar\omega) + J_{n-m}(qV_\omega/\hbar\omega)\right] J_{dark}(qV_{dc} + n\hbar\omega), \tag{1b}$$

where the static tunneling current (dark current) $J_{dark}$ flowing between two dissimilar metals can be calculated using the well-known Simmons' formula [26]-[27]. At infrared and visible wavelengths where $qV_\omega/\hbar\omega \ll 1$, the summation in Eq. (1) can be approximated by terms up to first order ($n = -1, 0, 1$). In this case, the relations between current density and electric field inside the MIM gap $E_\omega$ ( $V_\omega = \int_0^d E_\omega dz$ ) are derived as: $J_{dc} = J_{dark}(V_{dc}) + \sigma_0^{(2)}|E_\omega|^2$ and $J_\omega = \sigma_\omega^{(1)} E_\omega$, where the linear conductivity $\sigma_\omega^{(1)}$ and second-order quantum conductance $\sigma_0^{(2)}$, responsible for the optical rectification, are

$$\sigma_\omega^{(1)} = \frac{qd}{\hbar\omega} \frac{I_{dark}(V_{dc} + \hbar\omega/q) - I_{dark}(V_{dc} - \hbar\omega/q)}{2} \tag{2a}$$

$$\sigma_0^2 = \left(\frac{qd}{\hbar\omega}\right)^2 \frac{I_{dc}(V_{dc} + \hbar\omega/q) - 2I_{dc}(V_{dc}) + I_{dc}(V_{dc} - \hbar\omega/q)}{4} \tag{2b}$$

and *d* is the thickness of tunnel barrier. It has been experimentally observed and theoretically studied that the power dissipation due to quantum conductance may intrinsically limit the maximum field enhancement achievable in plasmonic nanostructures [1],[3],[7]-[9]. We note that contributions of higher-order harmonics (*m* > 2) are ignorable, since they are significantly attenuated in the proposed geometry.

Here we consider a HMM structure composed of stacked silver (Ag) and copper (Cu) thin-films with thicknesses $t_{metal}$, separated by 1-nm-thick niobium oxide (Nb$_2$O$_5$) insulating layers. The HMM slab is micromachined with a periodicity of 300 nm and 150-nm-wide air trenches. A plane wave with transverse-magnetic (TM) polarization is incident along the $-\hat{\mathbf{z}}$ direction. The trenched HMM structure in Fig. 1(a) is, in some sense, similar to a periodic array of waveguides with anisotropic cladding, for which a guided mode propagating along the waveguide axis may be tailored to have a zero group velocity, i.e., $v_g = \partial\omega/\partial\beta \approx 0$, at certain wavelengths, where $\omega$ is angular frequency and $\beta$ is the modal propagation constant. Based on Maxwell's equations and periodic boundary conditions, the dispersion relation between $\omega$ and $\beta$ can be derived based on the transverse resonance method [28] as

$$\tan\left[k'\frac{P-W}{2}\right] - \varepsilon_\perp \frac{k''}{k'} \tan\left[k''\frac{W}{2}\right] = 0, \tag{3}$$

where $k' = \sqrt{k_0^2 \varepsilon_\perp - (\varepsilon_\perp/\varepsilon_\parallel)\beta^2}$ (i.e., inside the HMM $k'^2/\varepsilon_\perp + \beta^2/\varepsilon_\parallel = k_0^2$), $k'' = \sqrt{\beta^2 - k_0^2}$, $k_0$ is the free space wave number, *P* is the period of unit cells, and *W* is the width of air slots. Provided that the periodicity of each MIM unit cell is subwavelength ($P \ll \lambda_0$), all diffraction orders, except for the zero-*th* mode, are evanescent.

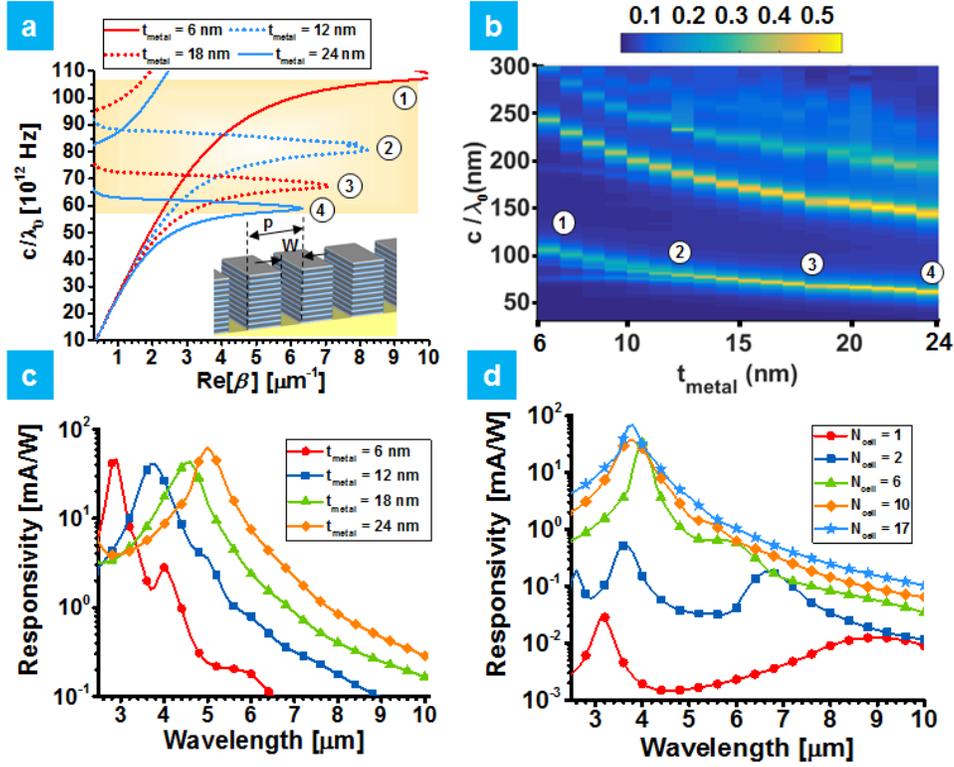

***Figure 2*** *(a) Dispersion diagram for a waveguide array made by carving a HMM substrate (P = 300 nm and W = 100 nm). (b) Contours of absorptance for the HMM in (a), varying the metal thickness ($t_{metal}$) and the operating wavelength; labels in (a) and (b) indicate the slow-wave modes and the corresponding absorption peaks. (c) Responsivity spectrum of the HMM in (a), varying the metal thickness (here the number of MIM unit cells $N_{cell}$ = 12). (d) is similar to (c), but varying the number of stacked MIM unit cells (here metal thickness $t_{metal}$ = 12 nm).*

Figure 2(a) shows the dispersion diagram of this HMM structure. In our calculation, we use realistic electronic and optical properties of materials extracted from previous experimental works [29], and considered the dissipative power due to quantum conductivity [Eq. (2a)]. It is seen from Fig. 2(a) that zero group velocity can be achieved at critical wavelengths, implying that the incident light is slowed down and trapped inside the air trenches, being significantly absorbed due to plasmon loss. Figure 2(b) shows the associated contour plot of absorptance as a function of metal thickness $t_{metal}$ [nm] and wavelength $\lambda$ [µm], calculated by full-wave simulation [30]; here 12 MIM heterojunctions are assumed and, for simplicity, the effect of substrate is neglected. We find excellent agreement between the wavelengths of maximum absorption in Fig. 2(b) and the zero-group-velocity points in the dispersion diagram [Fig. 2(a)]. We note that total HMM size is still subwavelength, since the large value of $\text{Re}[\beta]$ suggests that the incident light can be effectively absorbed in a long-wavelength range. Moreover, the absorption spectrum can be readily tailored by varying the volume fraction of metal, which determines the permittivity tensor elements of HMM.

The (photo)responsivity of this HMM-based structures can calculated as

$$\gamma = \frac{\sum_{i=1}^{2N-1} 2\int_{W/2}^{P/2} \sigma_0^{(2)} \left|\left(\mathbf{E}_\omega(x)\cdot\hat{\mathbf{z}}\right)\right|^2 dx}{\frac{\varepsilon_0 c}{2}\left|\mathbf{E}_{inc}\right|^2 P}, \qquad (4)$$

where $\mathbf{E}_{inc}$ is the electric field of the incident light, $\mathbf{E}_\omega(x)$ is the electric field inside the MIM gap, $\varepsilon_0$ and $c$ are permittivity and speed of light in free space. We note that $|\mathbf{E}_\omega|$ can be largely enhanced compared to $|\mathbf{E}_{inc}|$, due to the plasmon coupling across nanogaps [31], as discussed in [27]. Figure 2(c) shows the calculated responsivity spectrum for different metal thicknesses. It is seen that the peak responsivity can reach few tens of mA/W, with a scalable operating wavelength range that is determined by the volume fraction of metals. Figure 2(d) studies the influence of number of MIM unit cells ($N_{cell}$) on responsivity; here the metal thickness is fixed to 12 nm. As expected, the responsivity is increased by depositing more MIM unit cells. We find that a HMM consisting of a large number of MIM unit cells could have a responsivity up to 0.1

A/W level, without any external dc bias. A single layer MIM device (e.g. $N = 1$ for metasurface), however, shows poor responsivities compared with bulk metamaterials.

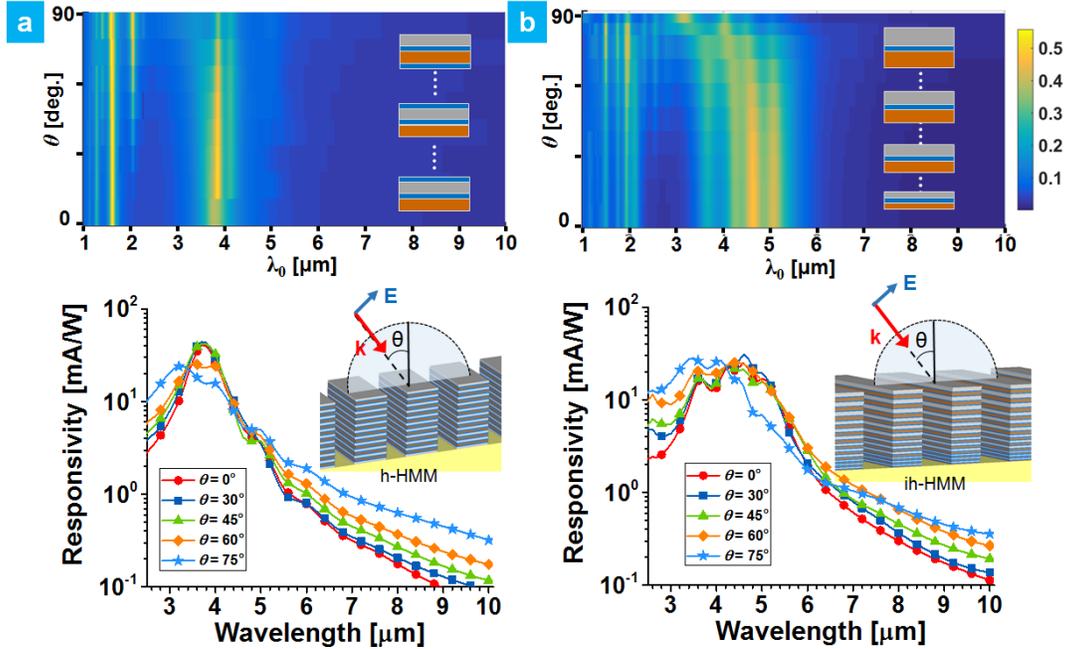

*Figure 3 (a) Contours of absorptance as a function of wavelength and incident angle of infrared for a homogeneous-HMM (h-HMM) device (top), and the corresponding responsivity spectrum (bottom). (b) is similar to (a), but for an inhomogeneous-HMM (ih-HMM) device.*

Interestingly, this HMM-based rectifying device may have the inherent advantage of being angle independent, because the operating wavelength simply depends on the eigenmode propagation inside each subwavelength HMM waveguide. Figure 3(a) shows the contour plot of absorptance as a function of operating wavelength and angle of incidence for the same devices in Fig. 2, with $t_m = 12$ nm and $N_{cell} = 12$ (top panel), and the associated responsivity spectrum for different angles of incidence (bottom panel). From Fig. 3(a), we find omnidirectional absorption at wavelength around 4 μm and a high zero-bias responsivity (~50 mA/W) at all angles. The linearly tapered HMM strips, such as the sawtooth geometry reported in Ref. [25], may be used to enhance the bandwidth of operation based on the superposition of multiple slow-wave modes. The gradient shape, however, would significantly increase the fabrication complexity and therefore reduce the device repeatability and uniformity. On the other hand, we know from Figs. 2(a) and 2(b) that the wavelength of slow light operation can be tuned by changing the volume fraction of metal. We propose here a simple and effective alternative design in Fig. 3(b), which no longer requires tapering the HMM strips. Instead, the superposition of several slow-wave modes is achieved by gradually varying the thickness of metal films along the light propagation direction. Figure 3(b) is similar to Fig. 3(a), but for an inhomogeneous HMM (ih-HMM) rectifying device with $t_{metal}$ varied from 24 nm, 18 nm, 12 nm to 6 nm (from top to bottom); here the total number of unit cells is the same as the homogeneous HMM (h-HMM) in Fig. 3(a). It is seen that broadband absorption, spanning from 3 um to 6 um, can be achieved due to the combination of multiple slow-wave modes [27]. The peak responsivity of ih-HMM is somewhat reduced compared to the h-HMM with same number of unit cells, due to the tradeoff between bandwidth and maximum absorption. The ih-HMM may achieve broadband, angle-independent, efficient mid-infrared rectification (a zero-bias responsivity ~30 mA/W), with maximum possible bandwidth highlighted in the dispersion diagram of Fig. 1(a).

Next, we compare the performance of h-HMM and ih-HMM devices with conventional infrared nano-rectennas, such as a broadband log-periodic tooth antenna (LPTA) or a resonant dipole antenna (RDA) [32] loaded with an MIM diode at the feedpoint. Figure 4(a) shows the responsivity for various infrared rectifying devices consisting of the same Ag-$Nb_2O_5$-Cu tunneling junction and a barrier thickness of 1 nm. The rectenna responsivity is calculated as the ratio of rectified current to the intercepted power (i.e., radiant power density multiplied by the antenna's effective aperture: $D\lambda_0^2/4\pi,$ where $D$ is the antenna directivity [32]).

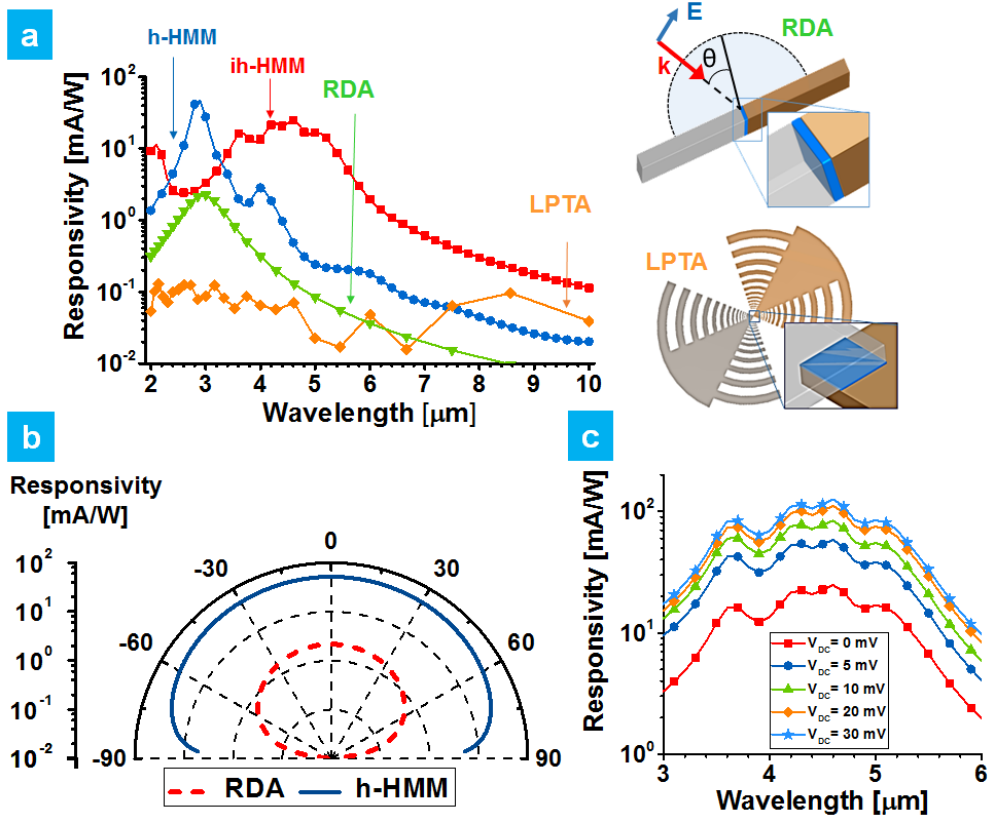

*Figure 4 - (a) Responsivity spectrum for different infrared rectifying devices made of the homogeneous-HMM (h-HMM), the inhomogeneous-HMM (ih-HMM), the resonant dipole antenna (RDA), and the broadband log period tooth antenna (LPTA); the right panel shows geometries of the RDA and LPTA, which are loaded with a MIM diode at their feedpoints. (b) Angular responses of h-HMM-based and RDA-based devices in a polar plot. (c) Responsivity spectrum of the ih-HMM-based device under a forward biasing.*

Here the MIM diode area is fixed to 100 nm × 100nm for LPTA and 30 nm × 30 nm for RDA, which is sufficiently small to avoid a parasitic cutoff frequency. From Fig. 4(a), it is seen that the HMM-based techniques can have much higher responsivities and broader bandwidth compared with lumped-element devices (0.01-1 mA/W). The responsivity values obtained for nano-rectennas are consistent with previous theoretical works [17]-[18]. For our nanopatterned HMMs, the improvement in responsivity may be attributed to the broadband and omnidirectional infrared absorption, combined with enhanced nonlinear optical rectification due to the interplay between plasmon coupling and photon-assisted tunneling within the MIM superlattices, making efficient usage of device area/density with massively parallel rectifying junctions. Figure 4(b) compares the angular response for the RDA-based and h-HMM-based devices. It is seen that the HMM-based device is rather insensitive to the angle of illumination, compared with conventional antenna-coupled designs. We note that the LPTA has a poor angular response due to the directive radiation pattern at short wavelengths. Figure 4(c) shows the responsivity of the broadband ih-HMM device in Fig. 3(b) under different forward-bias conditions. It is seen that the responsivity increases with forward biasing, and it can be greater than 0.1 A/W over a broad mid-infrared range, at a relatively small bias voltage.

In conclusion, we have put forward the concept and a realistic design of broadband, angle-independent infrared rectification based on the rainbow trapping effect in nanopatterned HMMs. Nonlinear optical rectification induced by photon-assisted tunneling in MIM-superlattice unit cells, when combined with concentrated optical fields in coupled plasmonic nanojunctions, can provide a 0.1 A/W-level responsivity at zero bias. This device does not require cryogenic cooling and hence enables ultrafast operation. We envision that this long-wavelength rectification method may be applied to several applications of interest at mid-infrared, including detection, imaging, sensing, communication, and energy harvesting.

# Supplementary Material for the Paper "Giant Photoresponsivity of Mid-Infrared Hyperbolic Metamaterials in the Quantum Regime"

## S1. Generalized Formula for the Electric Tunnel Effect in Metal-Insulator-Metal System

According to the Simmon's model [1]-[3], when a dc bias $V_{dc}$ is applied to the metal-insulator-metal (MIM) junction, the tunneling current density can be expressed as:

$$J_{dark}(V_{dc}) = \frac{4\pi m_0 q}{h^3} \int_0^\infty D(E_x) dE_x \int_{E_x}^\infty \left[ F_1(E) - F_2(E + qV_{dc}) \right] dE, \quad (S1)$$

where $q$ is the electron charge, $m_0$ is the electron mass, $h$ is the Plank constant, $F_i$ is the Fermi-Dirac distribution function of the $i$-th metal, $E$ and $E_x$ are respectively the total and longitudinal energy of electrons, $D$ is the electron transmission probability. The charge transport mechanism in the MIM junction depends on the difference between Fermi levels of two metals. When a dc bias is applied, currents can flow between two metals by means of the tunneling effect. At equilibrium, the well-known Simmons formula can describe the current-voltage behavior of this MIM junction [1]-[3]:

$$J_{dark}(V_{dc}) = J_{1\to 2} - J_{2\to 1} = J_{dark,0} \left[ \bar{\varphi} \exp(-A\sqrt{\bar{\varphi}}) - (\bar{\varphi} + qV_{dc}) \exp(-A\sqrt{\bar{\varphi} + qV_{dc}}) \right] \quad (S2)$$

where $J_{dark,0} = q/2\pi h(\beta \Delta s)^2$, $J_{1\to 2} = J_{dark,0} \left[ \bar{\varphi} \exp(-A\sqrt{\bar{\varphi}}) \right]$ is the current density flowing from metal 1 to metal 2, and $J_{2\to 1} = J_{dark,0} (\bar{\varphi} + qV_{dc}) \exp(-A\sqrt{\bar{\varphi} + qV_{dc}})$ is the current density flowing from metal 2 to metal 1, $A = 4\sqrt{2m_0}\pi\beta\Delta s/h$, $\beta$ is the correction factor defined in [1]-[3] as $\beta = 1 - [1/(8\bar{\varphi}^2 \Delta s)] \int_{s_1}^{s_2} [\varphi(x) - \bar{\varphi}]^2 dx$ ($s_1$ and $s_2$ are limits of barrier at Fermi level), $\bar{\varphi}$ and $\Delta s = s_2 - s_1$ are the mean potential barrier height relative to Fermi level and barrier width for tunneling electrons ($\Delta s$ could be different from the thickness of insulating film $s$).

In the reverse bias mode, of which the metal 1 with lower work function is negatively biased, the height of tunneling barrier between two metals, which takes into account the potential barrier lowering due to the image force, is given by

$$\varphi(x) = \varphi_1 + (\Delta\varphi - qV)\frac{x}{s} - \frac{q^2}{4\pi K\varepsilon_0 s}\left[\frac{x}{2} + \sum_{n=1}^{\infty}\frac{ns}{\left[(ns)^2 - x^2\right]} - \frac{1}{ns}\right]$$

$$\approx \varphi_1 + (\Delta\varphi - qV)\frac{x}{s} - 1.15\ln[2]\frac{q^2}{8\pi K\varepsilon_0}\frac{s}{x(s-x)}, \quad (S3)$$

where $\Delta\varphi = \varphi_2 - \varphi_1$, $\varphi_i = \Phi_i - \chi$ is the barrier heights at the interface between the $i$-th metal and oxide, $\Phi_i$ is the work function of the $i$-th metal, and $\chi$ is the electron affinity of oxide. From (S3), the mean barrier height $\overline{\varphi}$ is given by:

$$\overline{\varphi} = \frac{1}{\Delta s}\int_{s_1}^{s_2}\varphi(x)dx \approx \varphi_1 + \frac{s_2+s_1}{2s}(\Delta\varphi - qV) - 1.15\ln[2]\frac{q^2}{8\pi K\varepsilon_0 \Delta s}\ln\left[\frac{s_2(s-s_1)}{s_1(s-s_2)}\right] \quad (S4)$$

where limits of barrier at Fermi level $s_1$ and $s_2$ are given by the real roots of the cubic equation:

$$\varphi_1 + (\Delta\varphi - qV)\frac{x}{s} - 1.15\ln[2]\frac{q^2}{8\pi K\varepsilon_0}\frac{s}{x(s-x)}. \quad (S5)$$

From (S2) and (S4), the reverse-bias dark current density can be calculated.

In the forward bias mode, of which the metal 1 with lower work function is positively biased, the mean barrier height $\overline{\varphi}$ is given by:

$$\bar{\varphi} \approx \varphi_2 - \frac{s_2+s_1}{2s}(\Delta\varphi+qV) - 1.15\ln[2]\frac{q^2}{8\pi K\varepsilon_0 \Delta s}\ln\left[\frac{s_2(s-s_1)}{s_1(s-s_2)}\right] \quad (S6)$$

where $s_1$ and $s_2$ are given by the real roots of the cubic equation:

$$\varphi_2 + (\Delta\varphi+qV)\frac{x}{s} - 1.15\ln[2]\frac{q^2}{8\pi K\varepsilon_0}\frac{s}{x(s-x)}. \quad (S7)$$

From (S2) and (S6), the forward-bias dark current density can be calculated.

In a MIM structure, a transition to hopping or diffusive transport is difficult, since the field-induced breakdown occurs prior to reaching the bias necessary to bring about the change in mechanism, typical for quantum tunneling devices. When the applied bias is increased from zero to a high voltage, the carrier transport mechanism would transit from direct tunneling to field emission (Fowler Nordheim tunneling).

## S2. Eigenmode analysis of HMM-waveguide arrays

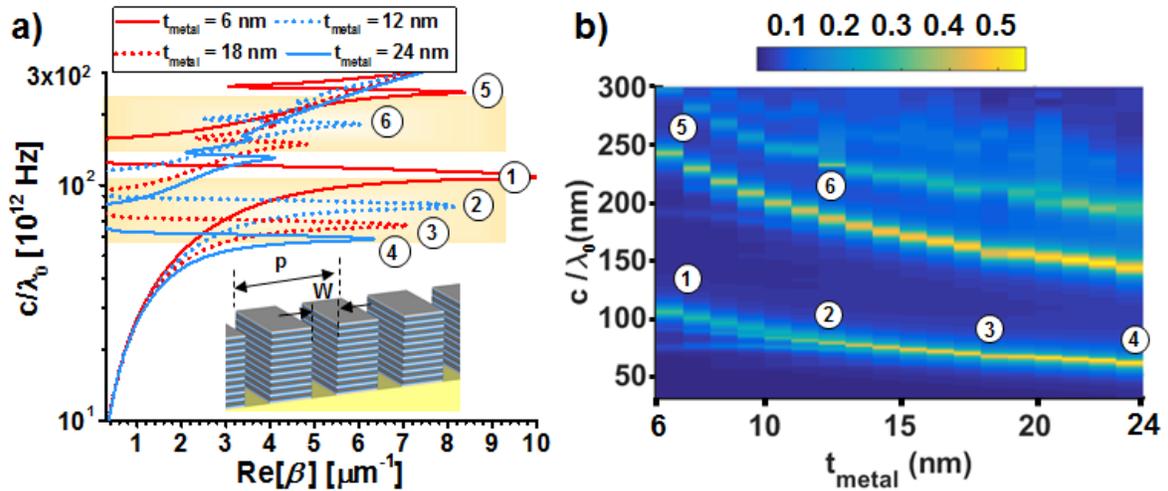

Figure S1: Dispersion diagram for (a) the HMM-waveguide array (P = 300 nm and W = 100 nm) made by carving a deposited HMM substrate; (b) Contours of absorptance for the HMM in

(a), varying the metal thickness (tmetal) and the operating wavelength; labels in (a) and (b) indicate the wavelengths of slow-wave modes and the corresponding absorption peaks.

Figure S1(a) reports the dispersion diagram of the nanopatterned HMM slab in Fig. 1(a) (also in the inset of Fig. S1(a)). Such geometry can be seen as an HMM-waveguide array. The MIM unit cells constituting the basis of HMMs can be designed to tame the spectrum of absorption (slow-wave modes). It is seen from Fig. S1(a) that there exists multiple slow-wave modes, which correspond to the absorption peaks in Figure S1(b). Further, slow-wave modes can be tuned over a wide range by varying the metal thickness.

## S3. Localization of light in MIM nanojunctions of HMM

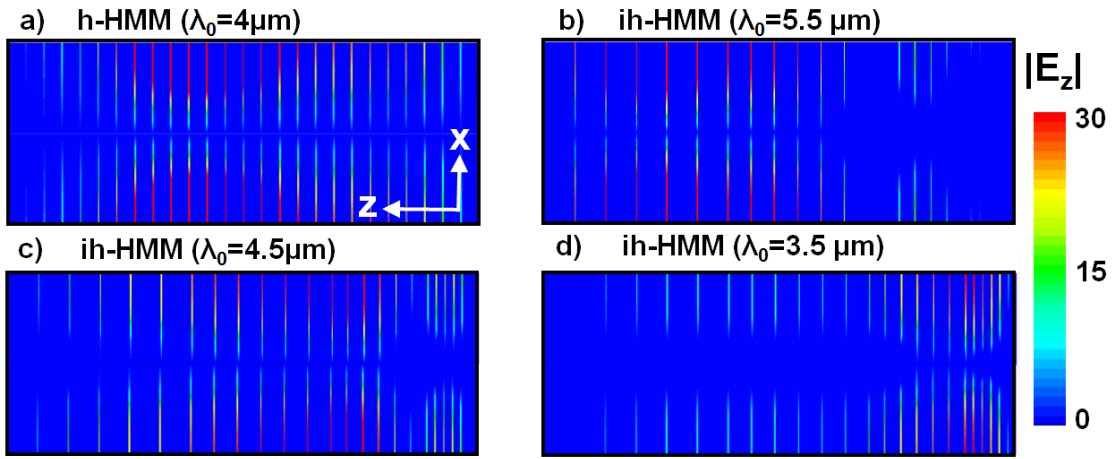

Figure S2: Electric field distribution ($E_z$) for (a) homogeneous hyperbolic-metamaterials (h-HMM) at its operating wavelength and (b)-(d) inhomogeneous hyperbolic metamaterial (ih-HMM) at different wavelengths.

Figure S2 shows the calculated contours of electric fields ($E_z$) for (a) a homogeneous-HMM (h-HMM) at its operating wavelength (4 μm) and (b)-(d) an inhomogeneous-HMM (ih-HMM) at different operating wavelengths. It is seen from Fig. S2(a) that electric fields are strongly localized inside the MIM tunnel junction and are directed parallel to the nanogap (perpendicular to the metal surface), which is expected to substantially enhance the tunneling-induced optical nonlinearities, such as the second-order nonlinear effect of optical rectification. From Fig. S2(b)-(d), it is seen that by tapering the unit cell dimension along the direction of light propagation, a broadband "rainbow trapping" effect can be achieved, of which different wavelengths of light are localized in different parts of the ih-HMM.

## S4. Tunneling-induced quantum conductivities

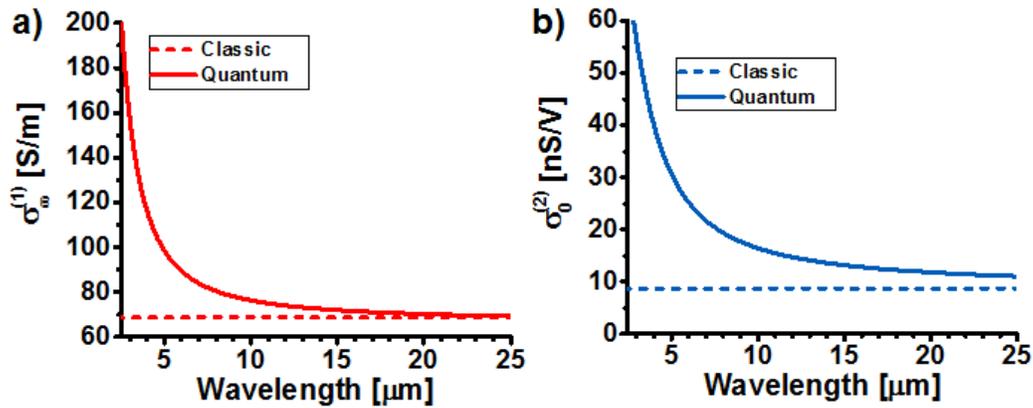

*Figure S3: (a) Linear and (b) second order quantum conductivity against frequency (solid: quantum mechanical model; dashed: classical model).*

Figure S3 (a) and (b) show the linear and second-order quantum conductivities against wavelength for an Ag-$Nb_2O_5$-Cu junction, which is calculated based on Eq. (2). The work functions of Ag and

Cu are $\Phi_1 = 4.26$ eV and $\Phi_2 = 4.7$ eV. The thickness, static relative permittivity, and electron affinity of Nb$_2$O$_5$ are $s = 1$ nm, $K = 25$ and $\chi = 4$ eV [4]-[7], respectively. It is seen that in the quantum regime (i.e. at high photon energies), the values of $\sigma_\omega^{(1)}$ and $\sigma_0^{(2)}$ are greater than what is expected from the classic small-signal treatment (dashed lines), valid only at long wavelengths (e.g. millimeter waves and microwaves).

## S5. Design of nano-rectennas

Here we designed two types of nano-rectennas: a broadband log-period-tooth antenna (LPTA) [8] and a narrowband resonant dipole antenna (RDA) [8]. The responsivity of a nano-rectenna is calculated as:

$$\gamma = \frac{\sigma_0^{(2)} |E_\omega|^2 A_{\text{diode}}}{P_{\text{inc}} A_{\text{eff}}} = \frac{4\pi \sigma_0^{(2)} |E_\omega|^2}{P_{\text{inc}} \, DIR} \frac{A_{\text{diode}}}{\lambda_0^2} \tag{S8}$$

where $E_\omega$ is the electric field across the nanogap of nano-rectenna, which is quite uniform and can be calculated using the equivalent nano-circuit model in Refs. [8]-[11]. In (S8), $P_{\text{inc}} A_{\text{eff}}$ represents the maximally possible power received by a nano-rectenna, where $A_{\text{eff}}$ is the effective aperture size of a nano-rectenna, which is related to the antenna directivity $D$ as: $D = 4\pi A_{\text{eff}} / \lambda_0^2$. [8] Figure S4(a) shows the calculated directivity and radiation pattern for RDA- and LPTA-shaped MIM diodes. For a subwavelength RDA, the radiation pattern basically follow a dipolar radiation over a wide wavelength range, and its directivity, according to the physical limit for a small scatter, is 1.5, rendering an effective aperture of $3\lambda_0^2 / 8\pi$. On the other hand, a LPTA with a larger physical size shows a higher directivity and bandwidth of operation. In the photodetection and/or photovoltaic applications, a directive beam pattern, however, implies a narrow angular operation range. Figure S4(b) compares the responsivity of RDA-based and LPTA-based mid-infrared rectifying devices. It is seen that the RDA-coupled MIM diode shows a larger responsivity than the LPTA-coupled one. This can be attributed to the improved conjugate matching condition.

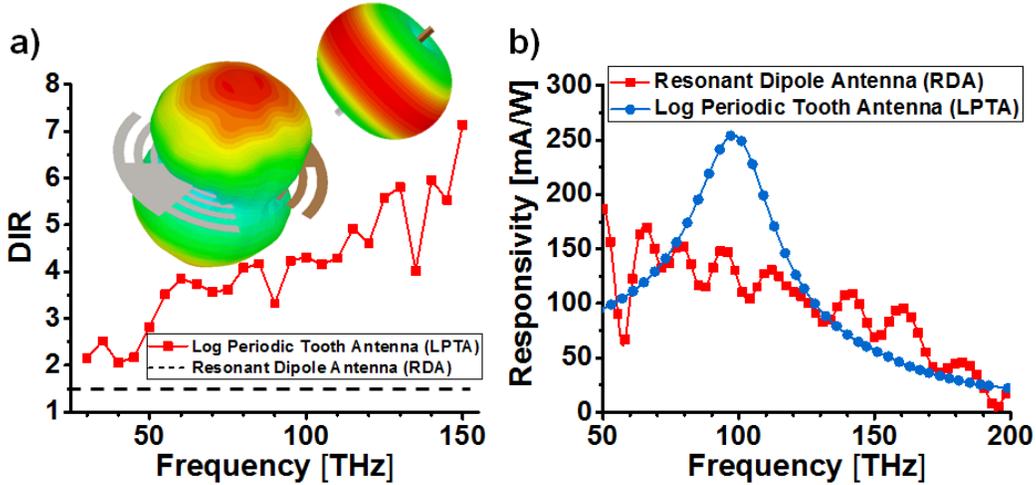

*Figure S4: (a) Directivity of RDA and LPTA loaded with a MIM diode. (b) Responsivity spectrum for RDA- and LPTA-coupled MIM diodes.*

Around the resonance of a RDA, the large inductive reactance of RDA can effectively cancel the capacitance of MIM diode, leading to a large field enhancement at the nanogap. However, the input impedance of a LTPA is almost non-dispersive and nearly real constant [8], which yields a lower field enhancement at the nanogap due to the parasitic cutoff.

**Reference**

[1] J. G. Simmons, *J. Appl. Phys.*, **34**, 1793 (1963).

[2] J. G. Simmons, *J. Appl. Phys.* **34**, 2581 (1963).